\RequirePackage[2020-02-02]{latexrelease}
\documentclass[%
 aip,
 amsmath,amssymb,
 reprint,%
]{revtex4-1}

\usepackage{graphicx}% Include figure files
\usepackage{bm}% bold math
\usepackage{amsmath,amssymb,amsfonts}
\usepackage{graphicx}
\usepackage{subfigure}
\usepackage[utf8]{inputenc}
\usepackage[T1]{fontenc}
\usepackage{mathptmx}
\usepackage{etoolbox}
\usepackage{subfigure}
\usepackage{afterpage}
\usepackage{xcolor}

\makeatletter
\def\@email#1#2{%
 \endgroup
 \patchcmd{\titleblock@produce}
  {\frontmatter@RRAPformat}
  {\frontmatter@RRAPformat{\produce@RRAP{*#1\href{mailto:#2}{#2}}}\frontmatter@RRAPformat}
  {}{}
}
\makeatother
\begin{document}

\preprint{AIP/123-QED}

\title{Coevolution of relationship and interaction in cooperative dynamical multiplex networks}

\author{Xiaojin Xiong}
 \affiliation{The College of Artificial Intelligence, Southwest University, No.2 Tiansheng Road, Beibei, Chongqing, 400715, China.}
 
\author{Ziyan Zeng}
 \affiliation{The College of Artificial Intelligence, Southwest University, No.2 Tiansheng Road, Beibei, Chongqing, 400715, China.}
  
\author{Minyu Feng}
 %\cormark[1]
 \email{myfeng@swu.edu.cn}
 \altaffiliation{Corresponding author}
 \affiliation{The College of Artificial Intelligence, Southwest University, No.2 Tiansheng Road, Beibei, Chongqing, 400715, China.}

\author{Attila Szolnoki}
\affiliation{Institute of Technical Physics and Materials Science, Centre for Energy Research, P.O. Box 49, H-1525 Budapest, Hungary.}

\date{\today}

\begin{abstract}
While actors in a population can interact with anyone else freely, social relations significantly influence our inclination towards particular individuals. The consequence of such interactions, however, may also form the intensity of our relations established earlier. These dynamical processes are captured via a coevolutionary model staged in multiplex networks with two distinct layers. In a so-called relationship layer the weights of edges among players may change in time as a consequence of games played in the alternative interaction layer. As an reasonable assumption, bilateral cooperation confirms while mutual defection weakens these weight factors. Importantly, the fitness of a player, which basically determines the success of a strategy imitation, depends not only on the payoff collected from interactions, but also on the individual relationship index calculated from the mentioned weight factors of related edges. Within the framework of weak prisoner's dilemma situation we explore the potential outcomes of the mentioned coevolutionary process where we assume different topologies for relationship layer. We find that higher average degree of the relationship graph is more beneficial to maintain cooperation in regular graphs, but the randomness of links could be a decisive factor in harsh situations. Surprisingly, a stronger coupling between relationship index and fitness discourage the evolution of cooperation by weakening the direct consequence of a strategy change. To complete our study we also monitor how the distribution of relationship index vary and detect a strong relation between its polarization and the general cooperation level.
\end{abstract}

\maketitle

\begin{quotation}
Social mechanisms are extensively studied to explain the formation of cooperators within networked populations. Interactions are characterized by a network, vertices represent agents engaged in evolutionary games, while edges can signify a variety of distinct connections between them. Moreover, multiplex dynamical networks provide a more suitable framework for a profound understanding of the simultaneous evolution of individual and collective states. Our study follows this research path where a multiplex structure is introduced including a relationship and an interaction layer. These layers affect each other simultaneously and our main focus is to follow their coevolution. To emphasize the importance of established relations we assume that the fitness of a player depends not only the payoff gained from interactions but also on the quality of personal relations. The latter is characterized by a relationship index which is subject to permanent change due to a coevolutionary process. We monitor how the distribution of this index changes by assuming different topologies of the relationship layer.
\end{quotation}

\section{Introduction}
In the development of human civilization, despite conflicts that have woven through history, cooperative behaviors have consistently manifested themselves. The exploration of cooperation in evolutionary dynamics constituted an enduring scholarly pursuit theoretically based on evolutionary game theory. Within the realm of strategy-making, social dilemmas symbolize the inherent tension between individual and collective interests within the population. A pivotal and frequently used model in evolutionary game theory is the prisoner's dilemma, which offers agents the binary choices of becoming cooperators or defectors.\cite{axelrod1981evolution}

The seminal work of Nowak and May has demonstrated that spatial setting of competitors, captured by a topological graph, could effectively enhance the evolution of cooperative behavior.\cite{nowak1992evolutionary} Subsequently, researchers concentrated on studying evolutionary games on various spatial topologies, including regular lattices,
\cite{nowak1993spatial,Rong2019} scale-free networks,\cite{santos2005scale}  small-world networks,\cite{ahmed_epjb00} and other complex networks.\cite{Abramson2000SocialGI} As an early summary of our understanding, Nowak distilled five mechanisms that drive the evolution of cooperation\cite{nowak2006five}. Through the creation of a dynamic effective network, it became evident that introducing autonomy promotes cooperative behavior.\cite{su2018promotion} As a further step, Civilini and his colleagues discussed an evolutionary game on hypergraphs where agents made choices between a risky option and a safe one.\cite{civilini2021evolutionary} Considering that agent payoffs are usually random in real situations, Zeng {\it et al.} argued that each agent's payoff follows a specific probability distribution with fixed expectations.\cite{zeng2022spatial} Building upon models with two players and two or three possible actions, Shi {\it et al.} investigated the dynamics of Q-learning within multi-agent systems. 

Coevolutionary rules introduced in evolutionary games capture the dynamic evolution of both strategies and the environment, hence offering a more accurate description. Initiated by Zimmermann {\it et al.}\cite{zimmermann2001cooperation} and influenced by the rapid advancements in network science, this field of research had since flourished as a promising avenue for addressing social dilemmas and promoting cooperation.\cite{Perc2009CoevolutionaryG} Coevolutionary rules could affect the interaction network,\cite{ebel2002evolutionary} the size of the network,\cite{szolnoki_epl08} agent selection,\cite{szolnoki2009emergence} or even individual mobility.\cite{xiao_zl_njp20} Moreover, evolutionary game theory in a networked population and its various extensions, like mixing game and multi-game, has been proven as an effective way to resolve the social dilemma.\cite{liu2019coevolution,chen_wm_pla22} A new spatial evolutionary game model explored cooperation by dynamically adjusting link weights among agents based on a comparison between their reputation and the average reputation of their nearest neighbors.\cite{2019Reputation} Mao {\it et al.} investigate how individuals' collective influence and strategy-updating time scales affect cooperation evolution.\cite{Mao2021Effect}

Various network models have been suggested to address the problems of collective behavior over the past decades, including multilayer networks,\cite{wang2012evolution,Boccaletti2014TheSA} temporal networks,\cite{li2020evolution} and higher-order networks.\cite{majhi2022dynamics} These novel network models served as the foundation for studying spatial evolutionary games. ``Multilayer networks'' was indeed a broad and general term, it could be further categorized into specific types,\cite{wang2015evolutionary} including multiplex networks,\cite{halu2013multiplex,gao2014single} interdependent networks,\cite{wang_z_srep13b,zhu2021information,su2023effects} and interconnected networks,\cite{wang2012evolution,huang2015cooperative,szolnoki_njp13} which provided more specific distinctions based on the nature of the network connections and interactions. For instance, G{\'o}mez-Garde{\~n}es {\it et al.} demonstrated that the multiplex structure of interdependent networks enhanced the resilience of cooperative behavior.\cite{gomez2012evolution} A recent work examined how various strategy-updating timescales impact the evolution of competing strategies, such as cooperation, defection, and extortion, in a double-layer lattice.\cite{Mao2023} Furthermore, multiplex networks were extensively used in network epidemic spreading models, enabling a more realistic exploration of the mechanisms involved in the interaction of multiple messages within social topics.\cite{li2019competition,feng2023impact}
Because of its practical importance multiplex networks frequently provide an adequate topological framework to study social dilemmas in evolutionary games. For example, an early work introduced the evolutionary game dynamics on structured populations where individuals take part in several layers of networks.\cite{GO2012Evolution} Additionally, Yu {\it et al.} delved into the impact of individuals' heterogeneous properties and the multilayer nature of networks on the evolution of cooperation.\cite{Yu2021Cooperation}  Furthermore, Hayashi {\it et al.} proposed a coevolutionary model in which strategy and layer selection are coupled hence each individual selects a layer and plays the social game with neighbors.\cite{Hayshi2016} Also in a coecolutionary framwork, Yang {\it et al.} introduced a model to study the impact of coordinating dynamic processes on the spread of strategies within evolving multilayer network.\cite{yang2019evolution}

In the real world, networks are often not only multilayered but also coevolutionary. Typically, agents usually conduct interpersonal relationships before engaging in evolutionary games. Moreover, during such games, agents do not necessarily interact with all their neighbors but may make selective choices.\cite{li2020autonomy} 
However, just a few published works paid attention to the interplay of relationship and neighbor selection based on a coevolutionary multiplex network in a social dynamics. This aspect, however, is a common phenomenon in real-life situations. To fill this gap, we assume that the dynamics governing the interactions among agents can be effectively captured by modeling them through distinct layers of a multiplex graph. Each agent has the opportunity to conduct relationships with other agents before interaction, leading to a greater willingness to interact with neighbors declared by the relationship graph. Throughout a coevolutionary process, following each round of games, the relationships between agents are influenced by their chosen strategies. More precisely, mutual cooperation strengthens an actual link, while simultaneous defection weakens it. As a consequence of bond evolution, the strengths of neighboring links also affect the individual fitness of focal player which is considered by an additional term via a relationship index, hence giving a novel perspective to the coevolutionary process.

This paper is organized as follows: we first introduce the applied social game and construct the proposed multiplex network. Section~\ref{sec:model} also explains the coevolutionary rule used during the simulation process. Our results, obtained on different relationship graphs, are summarized in Sec.~\ref{sec:results}. This section also discusses the connection between the time evolution of relationship index distribution and general cooperation level. We conclude with the summary of our findings and a discussion of their implications in Sec.~\ref{sec:conclusion}. Last, some future outlooks are also considered.

\section{Model}
\label{sec:model}
It is a ubiquitous phenomenon that there exists a complex interplay of mutual influence between the relationships and interactions among agents in social dilemma. In this section, we introduce a spatial coevolutionary model for the weak prisoner's dilemma (WPD) played on multiplex networks. We consider that each agent is willing to establish relationships with some other agents before playing the game, and the level of relationship between them could be different. To model this process, we construct a multiplex network that consists of two layers: a so-called relationship layer and an interaction layer. Our primary focus lies in the coevolution of agents' strategies and network structure, which is distinctly manifested in terms of cooperation density and interpersonal relationships among agents. We first define the WPD model characterizing the social dilemma, which is followed by the construction of evolutionary dynamics on multiplex networks. This also involves to elucidate the concepts of extended agents' fitness and the strategy update rule governing the microscopic dynamics.

\subsection{Weak prisoner's dilemma}
In a typical prisoner's dilemma, each agent has two distinct strategies: cooperation ($C$) or defection ($D$). During each round of interaction, both agents independently and simultaneously make their strategy choices. If both agents choose to cooperate, they each receive a reward denoted as $R$. Conversely, if both agents decide to defect, they both receive a payoff represented as $P$. However, when an agent chooses the defection strategy and encounters a neighbor who has chosen cooperation, the defector receives a higher payoff denoted as $T$, while the cooperative neighbor receives a lower payoff represented as $S$. Herein, the rank $T > R > P > S$ ensures the fundamental characteristics of the prisoner's dilemma game, in which the incentive for defection outweighs that for cooperation irrespective of the opponent's choice. For simplicity, but keeping the essence of the social frustration\cite{nowak1992evolutionary}, we employ the WPD parametrization by setting $R = 1$, $T = b$, and $S = P = 0$, where the parameter $b$ varies in the $1 \leq b \leq 2$ interval. 

\subsection{Multiplex networks}
\label{network}
To represent the above explained coevolutionary process faithfully, we construct a multiplex network where two layers are established. The two-layer structure allows for differentiation between the structural aspects of relationships (captured in the relationship layer) and the actual interactions and behaviors (captured in the interaction layer). The use of a multiplex network is justified when the interplay between structural relationships and behavioral interactions is crucial for understanding the system's dynamics.

In the relationship layer, each vertex represents an agent, and each edge denotes the relationship between two agents. Furthermore, each edge is characterized by a specific weight to indicate the reliability of the relationship between the involved agents. Initially, we assume that the weight of the edge is uniformly distributed in the range \([0, 1]\), which implies that any value of the edge weight between node $i$ and $j$ (denoted as \(W(i, j)\)) within the interval of 0 to 1 is equally likely.

For the interaction layer, each vertex has an equivalent in the relationship layer, i.e., represents the same agent. However, a key distinction lies in the fact that, in the interaction layer, agents probabilistically select neighbors for actual strategic interactions based on their structural positions in the relationship layer. In other words, the interaction layer is the arena where the actual dynamical evolution of strategic interactions occurs. At each time step, all agents can play the WPD against their chosen neighbors in the interaction layer. Furthermore, when two corresponding nodes are connected by an edge in the relationship layer, then there is a $p$ probability of game in the interaction layer. Conversely, if the corresponding two nodes are not linked in the relationship layer, a game is played with a probability of $(1-p)$ in the interaction layer. The latter process mimics the free will of players to interact with anybody else in the whole population.

As a crucial element of our model, we assume that game strategies also affect the $W$ weight values of edges in the relationship layer. If both neighboring agents choose cooperation then the weight of their edge will increase by $\epsilon$. If both agents defect, the weight of their edge will decrease by $\epsilon$. Otherwise, for unilateral cooperation the weight value remains intact. As a technical note, the weight value only changes to keep $W$ in the $[0,1]$. In a special case, when there is no association between the interacting agents, the relationship layer remains unchanged. In brief, the aforementioned rules can be expressed as 
\begin{equation}
W(i, j)=\left\{\begin{array}{ll}
[W(i, j)+\varepsilon]_{0}^{1}, & S_{i}=C \text {\ and\ } S_{j}=C \\

[W(i, j)-\varepsilon]_{0}^{1}, & S_{i}=D \text {\ and\ } S_{j}=D \\

W(i, j), & \text{otherwise}
\end{array}\right.
,
\end{equation}
where $S_{i}$ and $S_{j}$ are the game strategies of agent $i$ and $j$ respectively. $\epsilon$ denotes the change in the weight of related edge in the relationship layer, influenced by the strategies of involved agents. Here we use the operator $\left[\cdot\right]_0^1$ as 
\begin{equation}
    \left[a\right]_0^1=\left\{
    \begin{array}{cc}
       0, & a<0 \\
       a, & 0\leq a\leq 1 \\
       1, & a>1 
    \end{array}
    \right.
,
\end{equation}
which ensures that the weight values always remain in the $[0,1]$ interval.

\subsection{Strategy evolution}
\label{strategy}
In real-world social systems, fitness is often determined not only by payoffs but also encompasses various non-monetary factors such as reputation.\cite{fehr_n04,quan_j_pa21} But the quality of our relations could also be a benefit that can hardly be described by a payoff value. Motivated by this fact, we here introduce a relationship index as one of the ingredients of fitness. Accordingly, we define a relationship index $A_i$ of node $i$ as
\begin{equation}
    A_{i}=\Sigma_{\mathrm{j} \in \mathrm{V}} \mathrm{W}(i,j)\,,
\end{equation}
where $V$ is the set of all neighbors in the network, and each edge $e(i, j)$ in the relationship layer possesses a weight denoted as $W(i,j)$. It is important to stress that the introduced relationship index cannot be considered as a sort of reputation because its value also depends on the behavior of the partner. Instead, a particular $W$ value dedicated to a link characterizes the quality of their cumulative interaction. But of course, the value of $A$, which is calculated from all links of a player, may characterize the individual behavior an actor indirectly.

In agreement with common experience, we presume that agents who have more intensive relationships with other agents in the relationship layer exhibit higher fitness level. We postulate that fitness is proportional to both the payoff and the relationship index. To keep our model simple, we apply an additive approach and use a weight factor to adjust the relative significance of these two factors. Therefore we evaluate the fitness of agent $i$ according to 
\begin{equation}
    f_i (\Pi_i,A_i )=\Pi_i+mA_i\,,
\end{equation}
where $\Pi_i$ represents the payoff of node $i$, and $A_i$ is the relationship index to $i$. The weight factor $m$ is between 0 and 1, to adjust the relative importance of payoffs and the cooperation index. 

After accumulating the fitness in each round, all agents get a chance to update their strategies. Notably, each agent can only learn the strategies from their neighbors in the relationship layer. Hereby, an agent $i$ selects a neighbor $j$ in the relationship layer with the probability
\begin{equation}
    \centering
    s(i \rightarrow j)=\frac{W(i, j)}{A_{i}}
\end{equation}
for a potential strategy imitation.

Next, the probability that agent $i$ adopts the strategy of agent $j$ in the upcoming round of the game is determined by 
\begin{equation}
    \centering
    \Gamma\left(S_{i} \rightarrow S_{j}\right)=\frac{1}{1+e^{\frac{f_{j}\left(\Pi_{j}, A_{j}\right)-f_{i}\left(\Pi_{i}, A_{i}\right)}{K}}}
    ,
\end{equation}
where parameter $K$ describes the uncertainty of strategy imitation. When $K \rightarrow 0$, the strategy updating is definite. In that case, if $j$’s payoff is higher than $i$’s, $i$ adopts the strategy of $j$ with probability 1. On the contrary, if $j$’s payoff is lower, $i$ cannot adopt $j$’s strategy. In the other extreme case, when $K \rightarrow \infty$ , player $i$ updates the neighbor's strategy randomly independently of fitness values. In this work we set to $K=0.1$ which allows a strong selection with a certain noise.

\begin{figure}
	\centering 
	\includegraphics[width=\columnwidth]{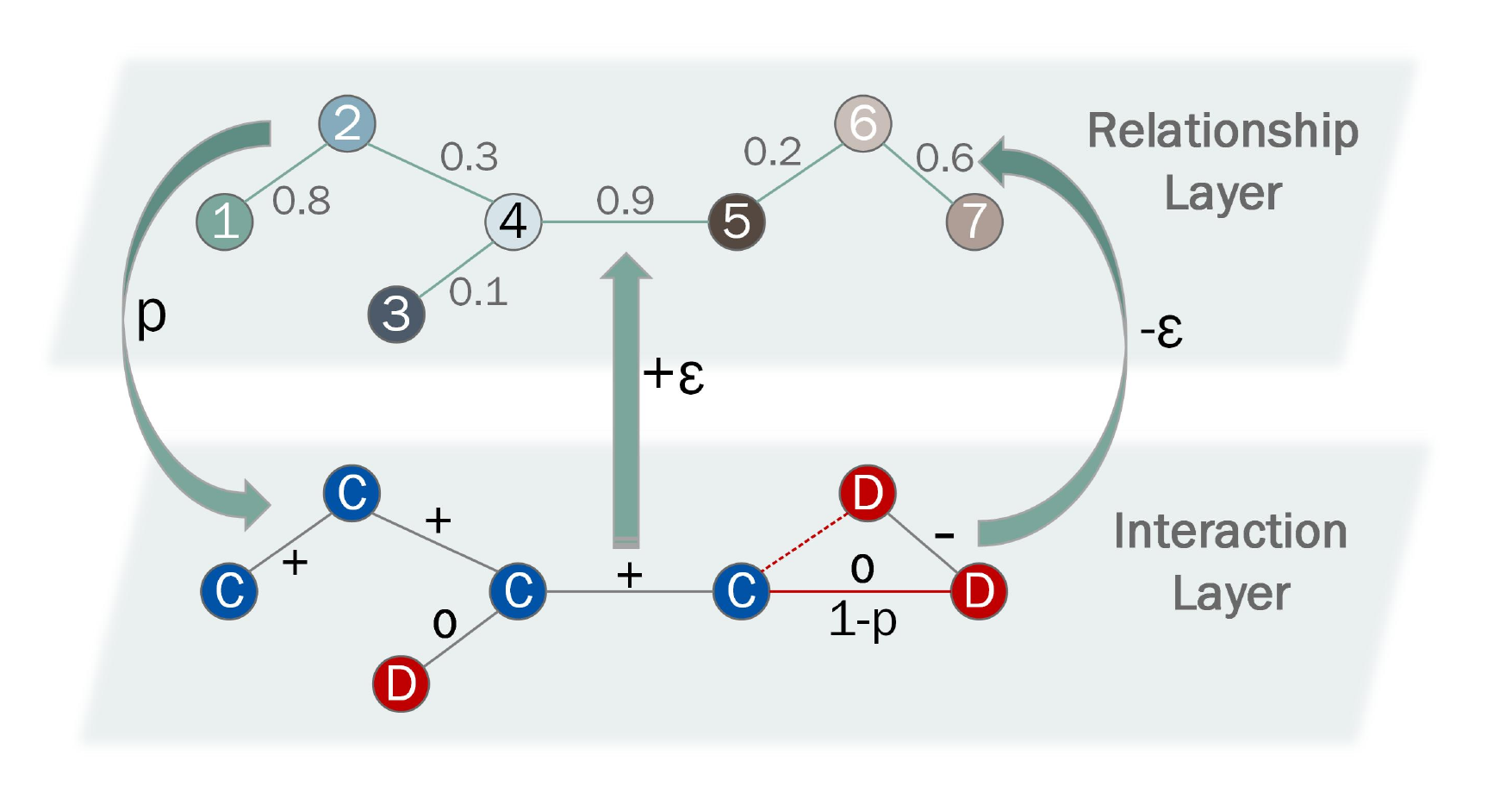}	
	\caption{\textbf{Coevolutionary game on multiplex networks.}
  The multiplex network consists of two layers, termed as ``relationship layer'', and ``interaction layer''. Each node within both layers corresponds to the same agent. In the relationship layer, weighted edges denote intensity of the association between agents. Edges in the alternative layer denote proper interactions among agents in the from of cooperation or defection. According to the coevolutionary rule, a link in the relationship layer will affect whether the game occurs in the interaction layer, and the outcomes of this game also effects the sate of the link in the relationship layer (color online).} 
	\label{fig1}
\end{figure}

\begin{figure*}  
    \centering

    \hspace{6mm}
    \subfigure[HL]
    {
    \includegraphics[scale=0.32]{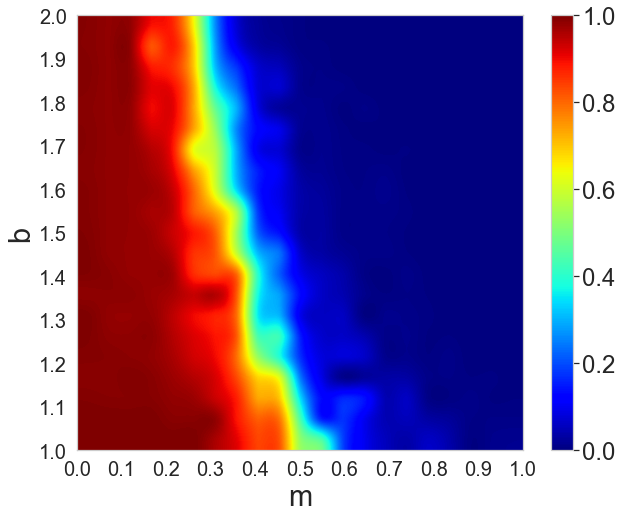}
    \label{fig2:3}
    }
    \hspace{2mm}
    \subfigure[SL]
    {
    \includegraphics[scale=0.32]{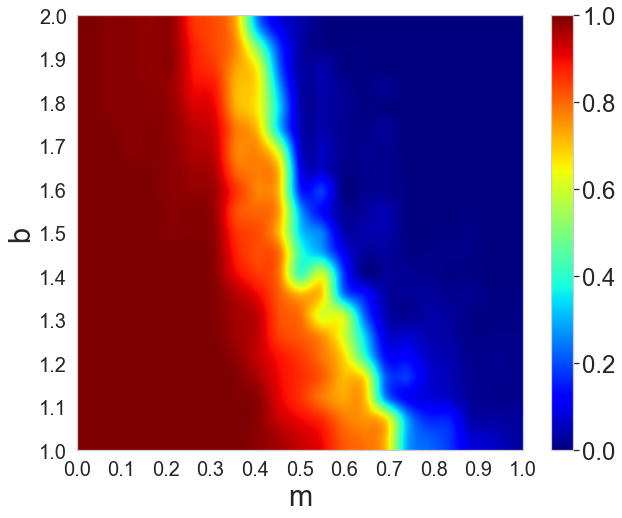}
    \label{fig2:4}
    }
    \hspace{-2mm}
    \subfigure[XL]
    {
    \includegraphics[scale=0.32]{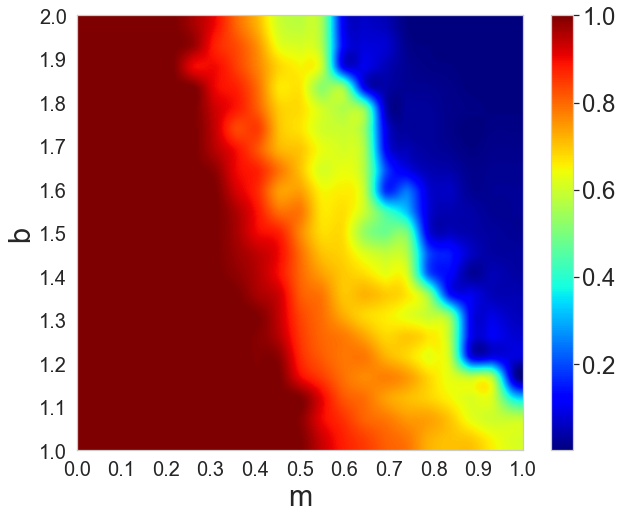}
    \label{fig2:6}
    }
    \hspace{2mm}
    \subfigure[WS]
    {
    \includegraphics[scale=0.32]{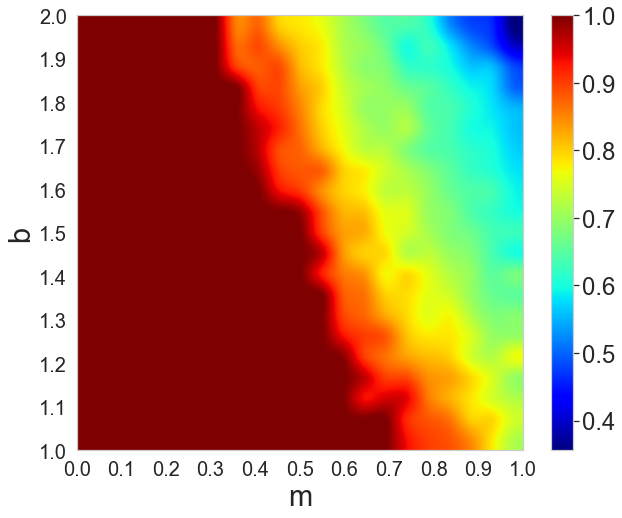}
    \label{fig2:WS}
    }
    \hspace{-10mm}
    \caption{\textbf{Cooperation heatmaps for four networks.} The $f_C$ density of cooperators is shown on the $b-m$ parameter plane at $p=0.9$ for honeycomb lattice (a), square lattice (b), hexagonal lattice (c) and WS small-world graph (d). The meaning of colors is explained in the right-hand side legend for each panel. It is generally valid that the cooperation level decays as we simultaneously increase the temptation value and the relative strength of relationship index in the fitness function. The stationary value of $f_C$ was averaged over 500 MC steps after 5500 relaxation steps in a graph containing $N=2500$ nodes (color online).}
    \label{fig2}
\end{figure*}

To summarize our model, Fig.~\ref{fig1} illustrates a typical scenario where agents in the relationship layer holds different relationships, quantified by the weights of edges. Each agent possesses an individual relationship index, influenced by the varying level of association with their neighbors, reflecting the diversity in the strength of interpersonal connections among individuals. The agents in the interaction layer correspond to the same as in the relationship layer and they all have two strategies to cooperate or to defect. It is noted that the neighbors of agent may not be the same in distinct layers, which is determined by parameter $p$, the possibility of relationship layer neighbors interacting in current games. For example, player 5 and 7 are not associated with each other in the relationship layer but they interact, hence build a link in the interaction layer with probability $(1-p)$. As mentioned in Sec.~\ref{network}, the neighbors' behavior also influence the intensity of their relationships. In the interaction layer, when both agents are cooperators (blue vertices), edges denoted as '$+$' means an increase in their corresponding relationship layers' edge weight by $\epsilon$. Conversely, when edges are marked as '$-$' for defectors (red vertices), it implies the intensity of their relationship decreases by $\epsilon$. Edges marked as '$0$' indicate that one agent cooperates while the other defects, hence their relationship remains unaffected. As illustrated in Sec.~\ref{strategy}, the behaviors of agents in the interaction layer evolve according to the defined strategy updating rule based on their extended fitness. In sum, the multiplex network structure provides an appropriate representation of the interplay between social structure and interactions, allowing for a more comprehensive understanding of the microscopic dynamics.

To reveal the possible role of the relationship layer on the system behavior we use different network topology. In particular, we apply honeycomb lattice (HL) with $k=3$ degree, square lattice (SL) having $k=4$, and hexagonal lattice (XL) where $k=6$. Besides, we also use Watts–Strogatz small-world network\cite{watts1998collective} (WS, $k=10$, with $p_r=0.5$ rewiring probability). For a proper comparison, the complete network contains $N=2500$ nodes in every cases. Initially, each agent is assigned randomly as a cooperator or a defector with equal probability. To reach the stationary state of the coevolutionary process we executed 5500 Monte Carlo (MC) steps and system-specific quantities, such as the fraction of cooperators or the mean of relationship index, are measured in the subsequent 500 MC steps. In this way the total length of simulation was 6000 MC steps for each parameter values. To ensure the accuracy of the experiments, our results represent the average of 10 repetitions for each experiment. The simulation procedures are implemented by using Python 3.9.

\section{Results}
\label{sec:results}
In this section, we present our observations about the coevolution of cooperation level and relationship index by illustrating how cooperative behaviors vary with game parameters. 
\subsection{Evolution of cooperation in coupled multiplex graphs}

We first present how the cooperation level changes by varying the main model parameters which are the $b$ temptation to defect and the $m$ relative weight factor of relationship index in the extended fitness function. To gain a generally valid observation about the system behavior we present results simultaneously obtained for different graphs of relationship layer. The results are summarized in Fig.~\ref{fig2} where we set $p=0.9$ for all cases. This $p$ value practically means that players interact with their neighbors mostly but there is a small chance to play a game with other players, too, with whom there is no permanent relationship. Across all four network types, cooperation density decreases as either parameter, $b$ or $m$ increases. Within the explored parameter range, both pure cooperation and pure defection regions are observed. Generally, smaller values of $b$ and $m$ lead to a higher prevalence of cooperation, while larger values are associated with an increased presence of defection. Furthermore, the transition from pure cooperation to pure defection exhibits a left-skewed trend for all network types.

Staying at the lattice structures, it is noted that enhancing the degree has been observed to favor cooperation generally. Both the HL and the SL cases exhibit a sharp decline in cooperation density, forming a distinct transition line on the parameter plane. While the XL and the WS topologies provide more chance for the coexistence of cooperation and defection, thereby suppressing the occurrence of pure defection in the population. Within the parameter range we investigated, defective strategy predominates in the HL case. Conversely, as to WS, the space for pure defection is greatly restricted, existing only when both $b$ and $m$ reach their maximum values.

\begin{figure*}  
    \centering
    \subfigure[]
    {
    \includegraphics[scale=0.27]{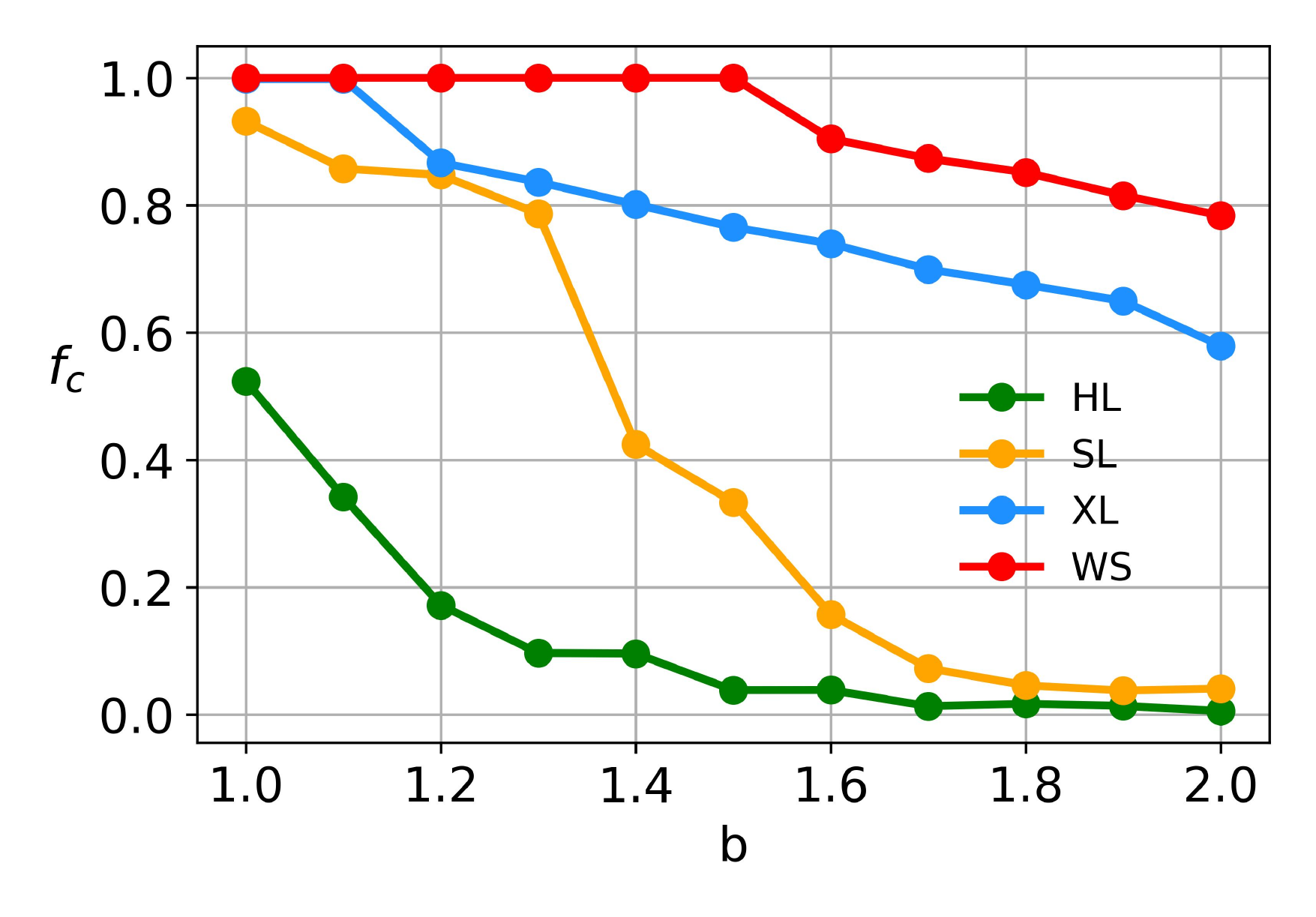}
    \label{fig3:1}
    }
    \hspace{8mm}
    \subfigure[]
    {
    \includegraphics[scale=0.27]{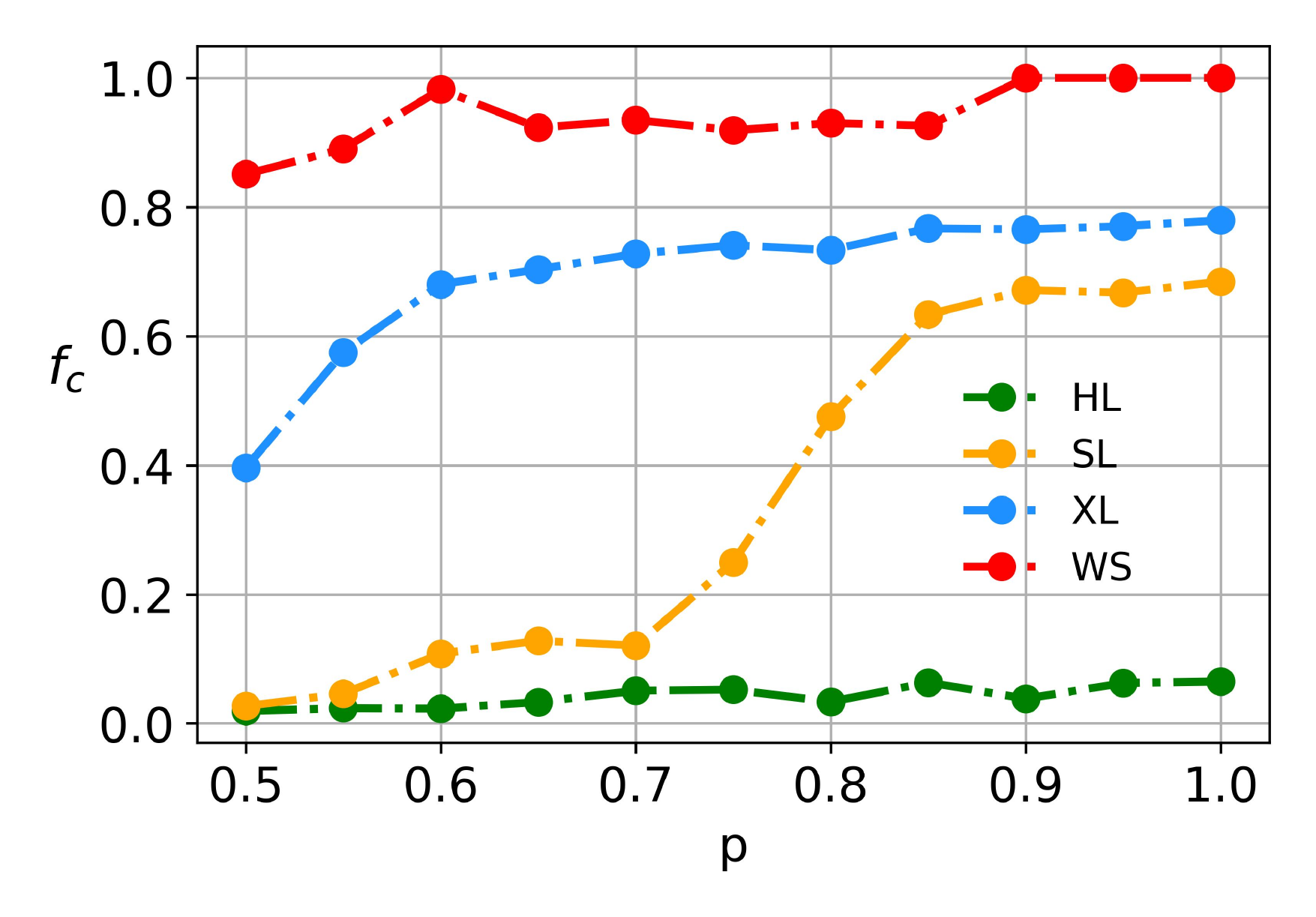}
    \label{fig3:2}
    }

    \caption{\textbf{Cooperation level in dependence on parameter $b$ and $p$ for four networks.} Panel~(a): $f_c$ as a function of $b$ at fixed $m=0.50$ and $p=0.90$ values. Panel~(b): $f_c$ as a function of $p$ at fixed $b=1.50$ and $m=0.50$ (b). Curves with different colors represent different topology of relationship layer as indicated in the legend: HL (honeycomb), SL  (square), XL (hexagonal lattice), and WS (small-world).}
    \label{fig3}
\end{figure*}

Next, as shown in Fig.~\ref{fig3:1}, we investigate the variation of $f_c$ with respect to $b$ when $m=0.50$ is fixed. The results here corroborate the aforementioned observations, namely, a higher degree in the three regular lattice networks (HL, SL, XL) provides more favorable conditions for the evolution of cooperation. Within the range $1 < b < 2$, the cooperation density for HL case consistently remains below 0.60 and rapidly declines towards 0 as $b$ is decreased. Square lattice topology offers a significantly different system behavior where there is a steep decline in the cooperation density between the range of $b$ values from 1.30 to 1.60, continuously diminishing from an initial value close to 1.00 and approaching 0 across the remaining parameter range. Moreover, among the four network types we investigated, it is noteworthy that under these parameter conditions, the steepest decline of $f_c$ occurs within the SL case. The $f_c$ function of XL topology, on the other hand, maintains a steady decline after $b$ exceeds 1.10 but consistently involves a significant portion of cooperators. For the WS case, the cooperation density remains maximal at around $b=1.0$, with a gradual decrease observed only after $b$ surpasses 1.50, yet it remains consistently above 0.75.

As expected, larger temptation involves lower cooperation level, but the way how $f_C$ decays could be qualitatively diverse for different topologies. The system behavior is the most sensitive on $b$ for square lattice and the temptation has just less importance for those graphs where the average degree is high.

Our findings align with conventional models, indicate that increasing the temptation to defect ($b$) consistently leads to a progressive decline in the population of cooperators. Notably, the distinctive features of our model highlight the importance of parameter ($m$) which governs the relative significance of the relationship index within extended fitness. In particular, we observe that elevating the value of $m$ consistently results in the decay of cooperation across all network structures employed in our study. In essence, a higher significance assigned to the relationship index in the fitness formula negatively affects cooperation.

In our previous simulations, we maintain a fixed probability ($p=0.90$) for establishing connections within the interaction layer according to the relationship layer. However, to investigate the potential impact of $p$, which determines how strictly an agent interacts with neighbors defined by the relationship layer, we conduct simulations by systematically changing its value at fixed $b=1.50$ and $m=0.50$. As shown in Fig.~\ref{fig3:2}, the variation in $p$ results in largely different system reaction at specific topologies. A relatively insignificant impact on cooperation behavior can be detected for honeycomb lattice and small-world graphs. Notably, under these parameter conditions, cooperation level in the HL consistently remains close to 0, regardless of the specific $p$. However, for square and hexagonal lattices, the stationary cooperation densities exhibit a nearly continuous growth as $p$ increases. The most visible difference between lattice structures is that square lattice shows the most sudden growth in cooperation density within the $0.7 < p < 0.9$ interval, while hexagonal lattice depicts the largest cooperation growth for lower $0.5 < p < 0.6$ values. Based on the comparison, we can conclude that square lattice case shows the most pronounced improvement of cooperation over the entire $0.5 < p < 1$ interval.

\subsection{Clustering driven by relationship layer of square lattice}
\begin{figure*}                                       
    \centering
    \hspace{-2mm}
    \subfigure[]
    {
    \includegraphics[scale=0.3]{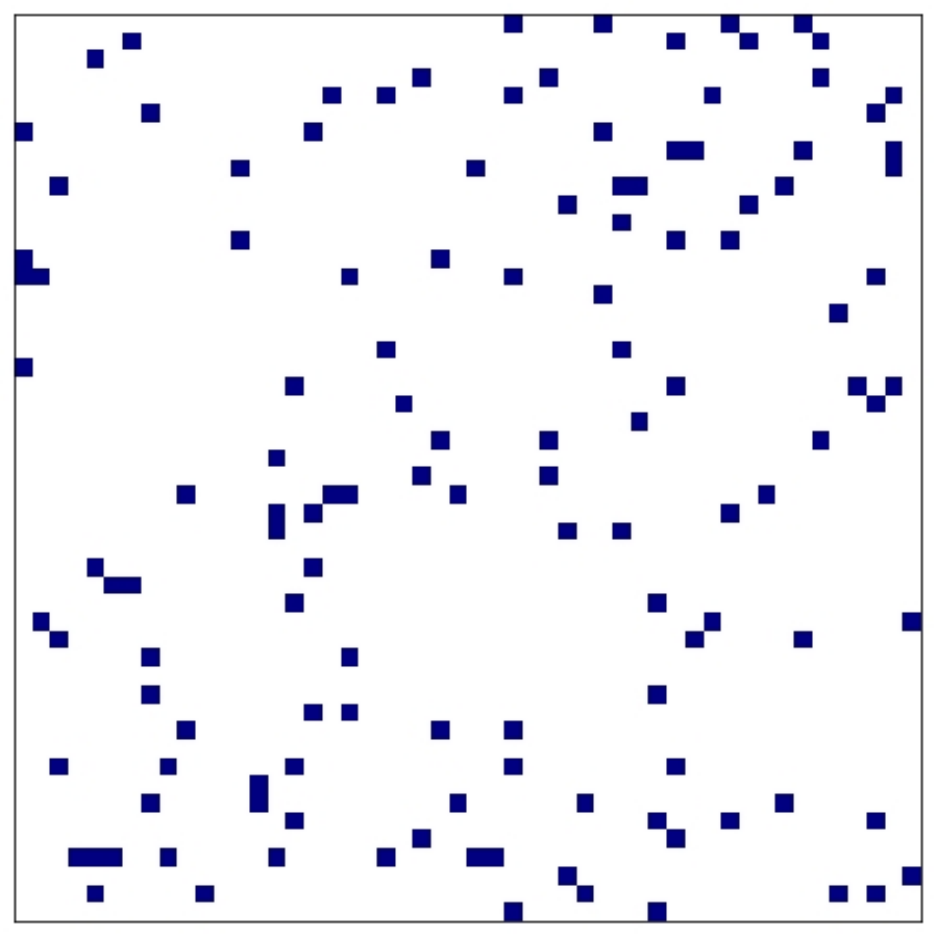}
    \label{fig-1}
    }
    \hspace{7mm}
    \subfigure[]
    {
    \includegraphics[scale=0.29]{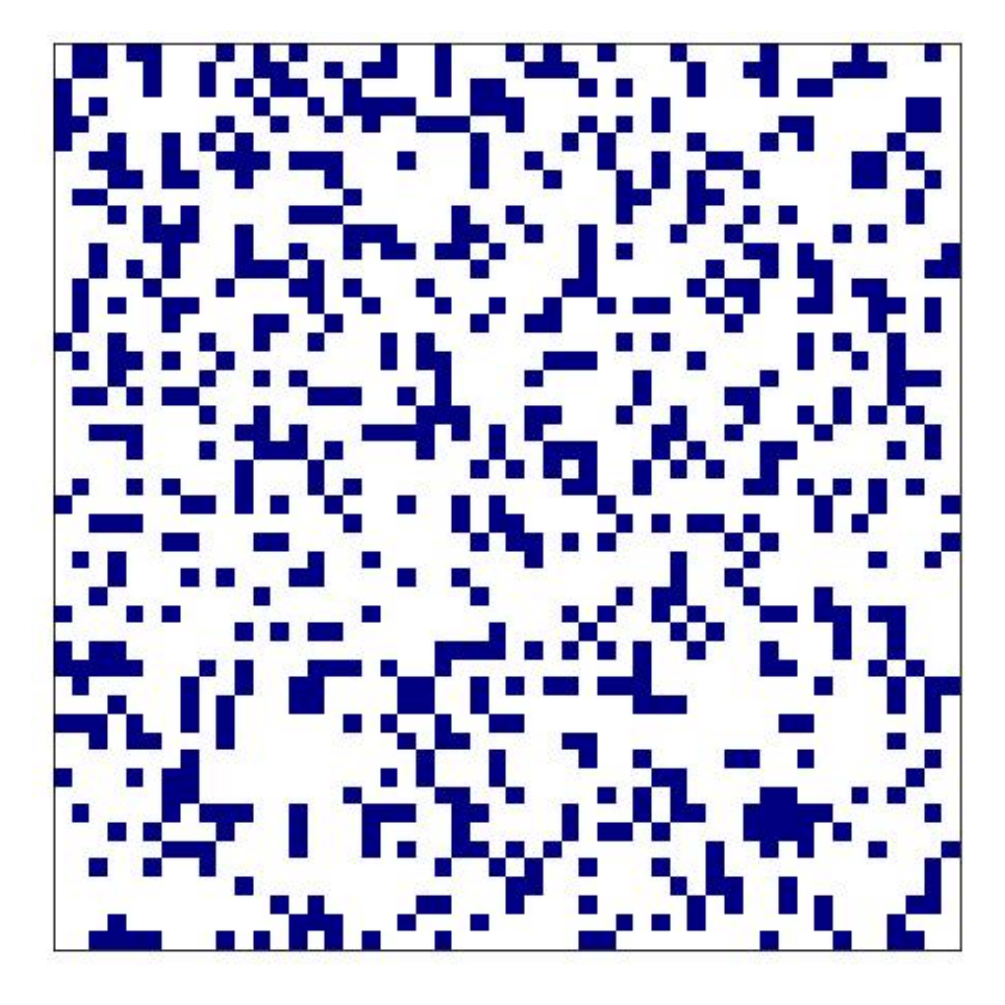}
    \label{fig-2}
    }
    \hspace{9mm}
    \subfigure[]
    {
    \includegraphics[scale=0.3]{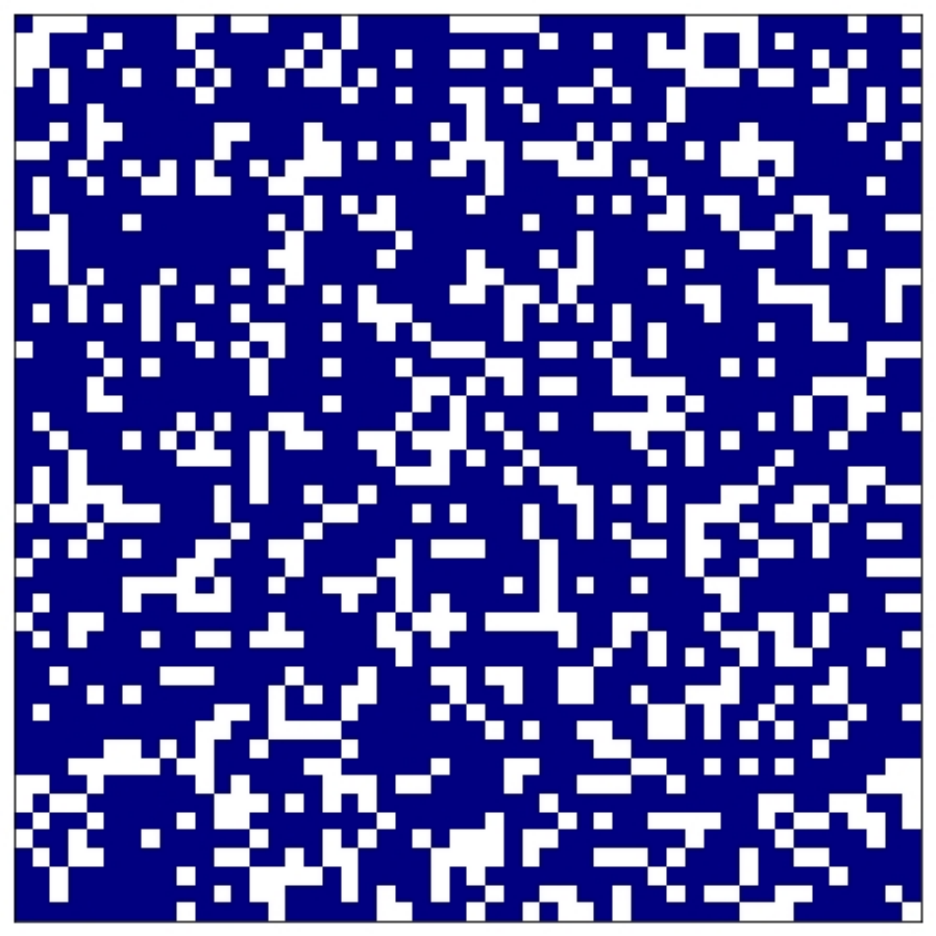}
    \label{fig-3}
    }
    \hspace{5mm}
    \subfigure[]
    {
    \includegraphics[scale=0.3]{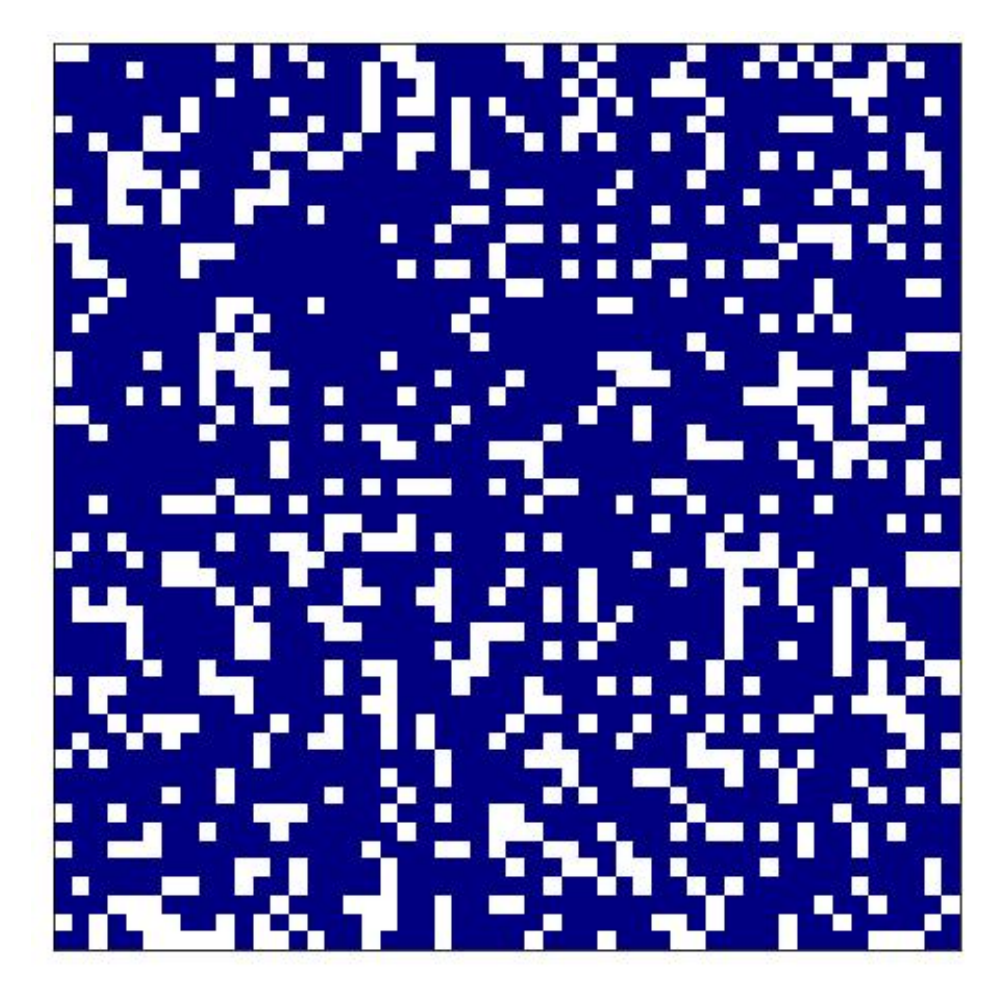}
    \label{fig-4}
    }
    \hspace{5mm}
    \subfigure[]
    {
    \includegraphics[scale=0.3]{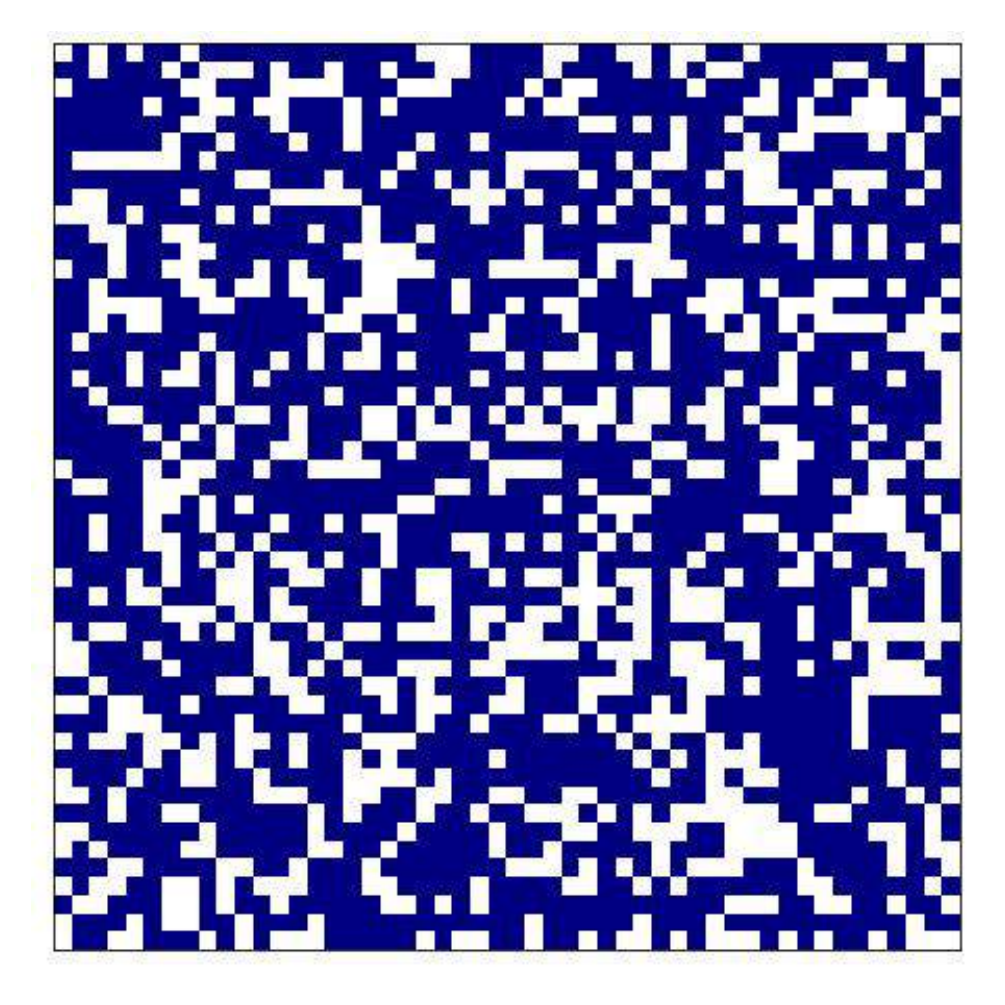}
    \label{fig-5}
    }
    \hspace{5mm}
    \subfigure[]
    {
    \includegraphics[scale=0.3]{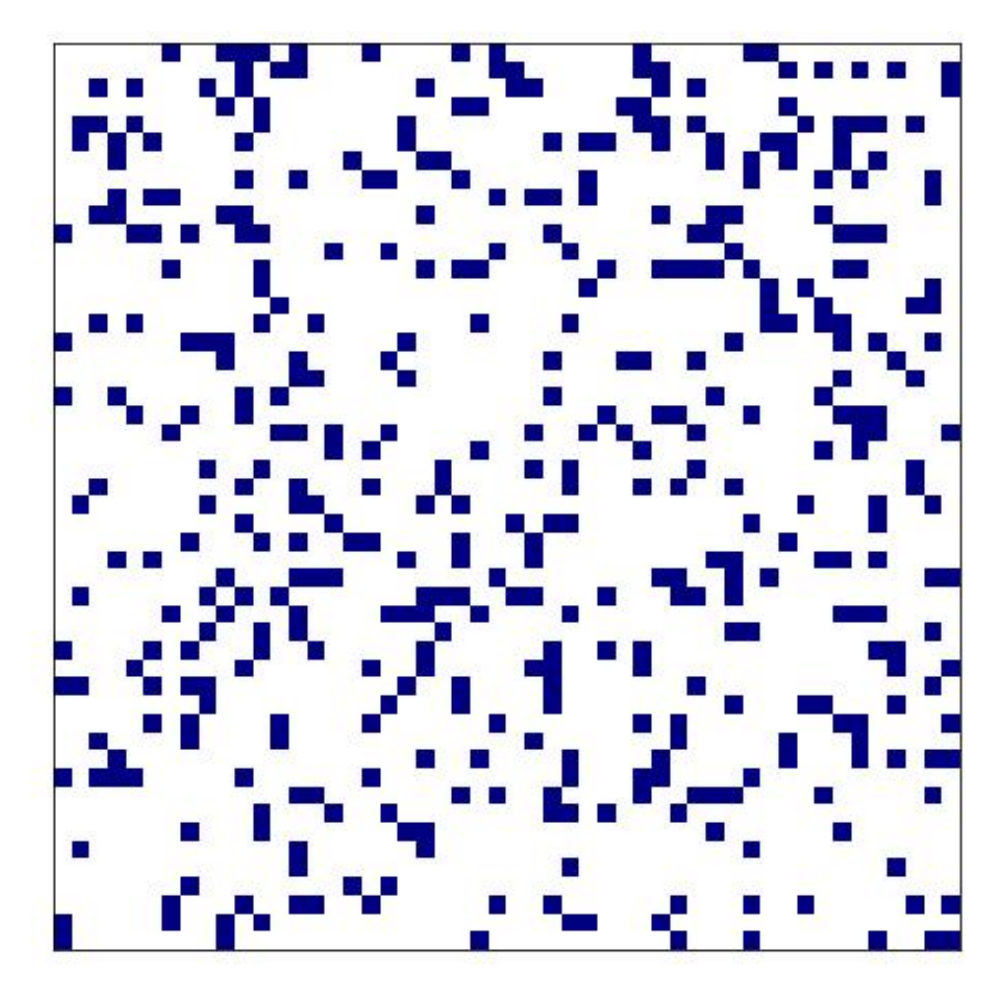}
    \label{fig-6}
    }  
    
    \caption{\textbf{Clustering of cooperators when using square lattice topology in relationship layer.}
     On the upper row we use $b=1.5$ and $m=0.5$ for all cases at different $p$ values, which are $p=0.5$ (a), 0.75 (b), and 1 (c). On the lower row we apply $p=0.9$ and $m=0.5$ at various temptation levels. They are $b=1.3$ (d), 1.5 (e), and 1.7 (f). Cooperators are marked by blue while defectors are denoted by white color in a lattice containing 2500 players. The upper low indicates that we need a large $p$ value to spatial reciprocity work. For moderate $p$ values large cooperator islands cannot form. The lower row illustrates that the cooperation level remains reasonable even at $b = 1.5$ and the major decay of cooperation happens for larger $b$ values. Below this parameter condition cooperators can maintain their tight formation but above this the change is significant. This particular behavior, exclusively characterizes square lattice, separates other cases observed for alternative topologies where the temptation value has less critical role on the cooperation level.}
     \label{fig4}
\end{figure*}

As we stressed, the utmost sensitivity in system behavior becomes apparent when the relational layer assumes the structure of a square lattice. This topology represents a kind of tipping point where nodes have relatively small, but not too small degree. In this way the topology can reveal the impacts of model parameters delicately which remain hidden when a general player has too few or too many neighbors. The spatial strategy organization for this special case is shown in Fig.~\ref{fig4} where we present the stationary distributions for some typical model parameters.

We first consider the temptation to defection as $b = 1.5$ and the relative fitness weight parameter as $m = 0.5$ on the upper row of Fig.~\ref{fig4}. As we increase the parameter $p$ from 0.50 to 1 we can see that the portion of cooperators grows from 0.053 (a) to 0.684 (c). But the increment is not continuous because even at $p=0.75$ the portion of cooperators is only 0.272 (b). These panels illustrate nicely that the clustering of cooperators due to network reciprocity can only work for large $p$ values. In particular, we cannot really detect compact blue islands in panel (a) and in panel (b) because the modest $p$, or in other words the relatively high $1-p$ probability, allows player to establish temporary interactions with distant players. In this way network reciprocity cannot work efficiently. For alternative topologies where the average degree is higher, such as for XL and WS cases, this effect is less significant because the crowded neighborhood suppress it.

On the lower row of Fig.~\ref{fig4}, we systematically adjusted the parameter $b$ to values of 1.30 (d), 1.50 (e), and 1.70 (f) at fixed $p=0.9$ and $m=0.5$ values. As expected, by increasing $b$ the portion of cooperators decays from 0.771 (d) to 0.177 (f). However, this decline takes a noteworthy turn above $b=1.50$. It is important to highlight that even at a relatively high value of $b = 1.50$ in panel (e), a substantial cooperator density of 0.619 remains significant. This phenomenon illustrates that for large $p$ values network reciprocity, supported by the extended fitness function, could be efficient no matter the temptation to defect is relevant.

Numerous prior studies have underscored the pivotal role of network structures in fostering cooperative clusters and subsequently shielding against invasions by defectors. This phenomenon is driven by the necessity for each agent to adopt a cooperative strategy when interacting with their neighbors, ensuring the sustenance of substantial mutual payoffs. 
As it is illustrated in the spatial strategy distributions of Fig.~\ref{fig4}, cooperator players cannot maintain large clusters anymore at relatively low $p$ and significantly high $b$ values, but become isolated, hence they have no chance to utilize the positive consequence of network reciprocity. Additionally, when an agent receives a meager payoff, notably smaller than that of its defector neighbors, it leads to the rapid disintegration of the cooperative cluster, presenting a challenge in terms of reconstitution. When we extended the fitness function by considering individual relationship index, our original motivation was to explore whether this additional term can improve the conditions for cooperation. Interestingly enough, to involve the local state of relationship has the opposite effect because it covers the pure consequences asocial behavior. While an actual $D-D$ bonds gives nothing to payoff, according to WPD parametrization, the $W$ value of this bond could be significant because it can cumulate a better $C-D$ state from the past. In this way the feedback of strategy choice cannot be as straightforward as for the case when fitness is determined purely by payoff values.

\begin{figure}
	\centering 
	\includegraphics[scale=0.3]{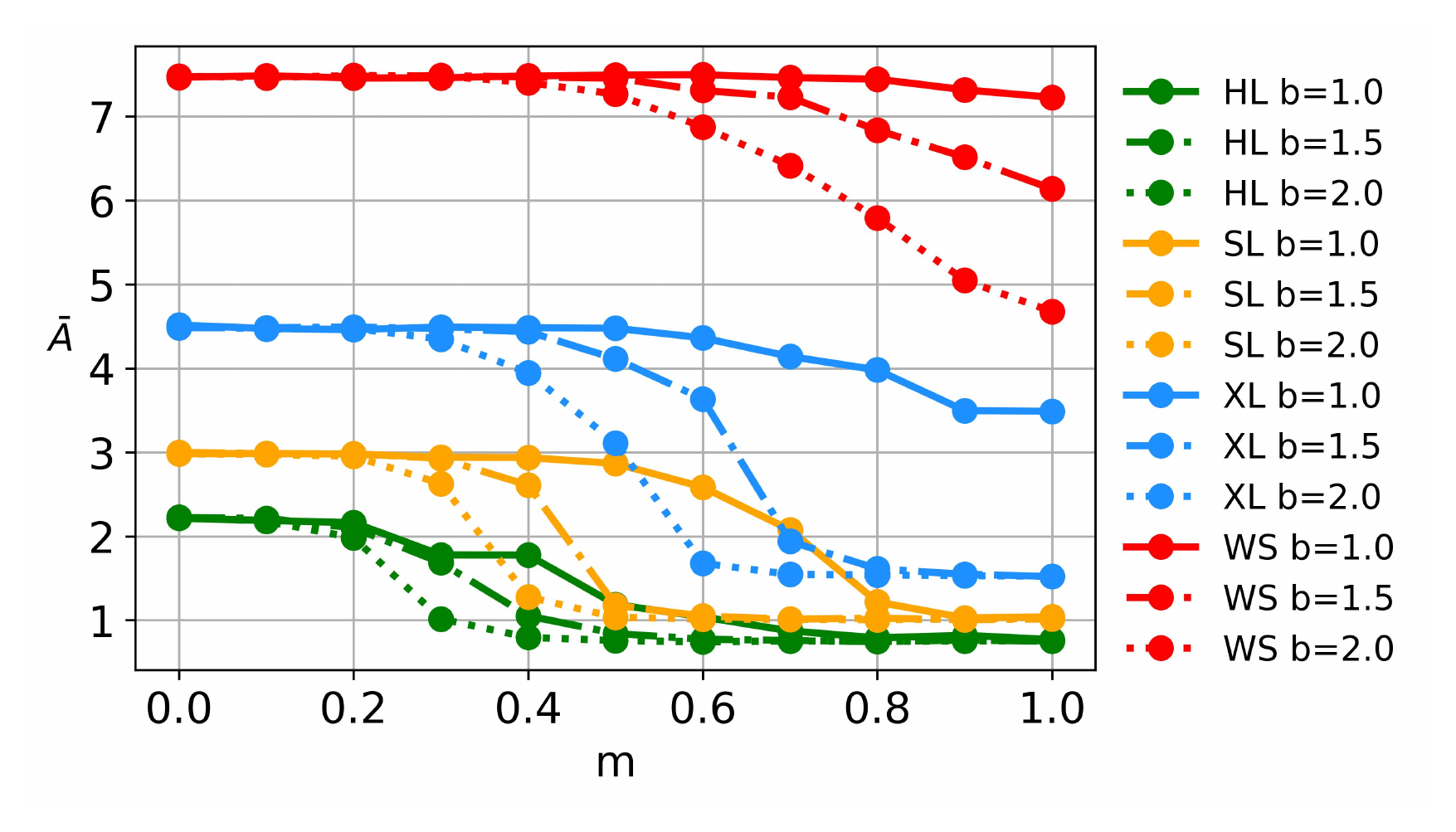}
	\caption{\textbf{The mean of relationship index $(\bar{A})$ plot in dependence on $m$ and $b$ for four networks.}
We demonstrate the influence of parameter $m$ on the mean of the relationship index for four distinct network structures at $p=0.9$. Different colors indicate different network type as indicated in the legend. They are honeycomb (HL), square lattice (SL), hexagonal lattice (XL) and small-world graph (WS). Solid, dash-dot, and dashed lines respectively correspond to $b=1.0$, 1.5, and 2.0 values.}
	\label{fig5}
\end{figure}

\begin{figure*}                                       
    \centering
    %\hspace{-12mm}
    \subfigure[HL]
    {
    \includegraphics[scale=0.27]{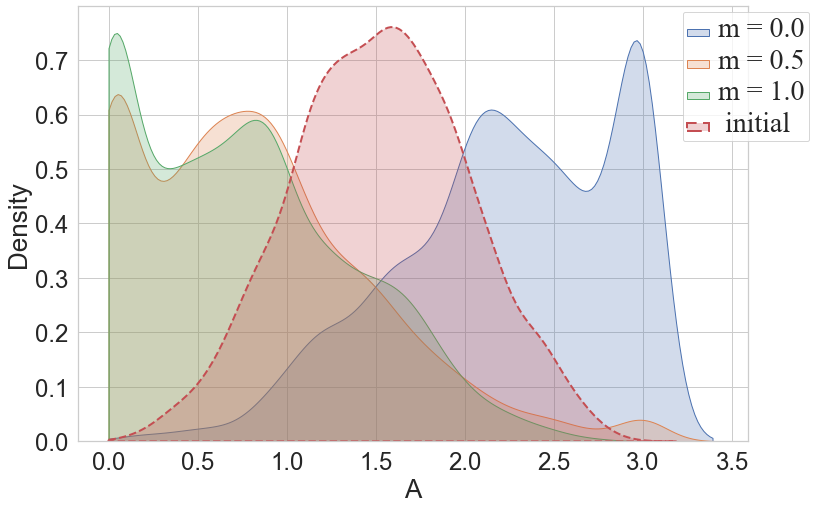}
    \label{fig:3}
    }
    \hspace{-2mm}
    \subfigure[SL]
    {
    \includegraphics[scale=0.27]{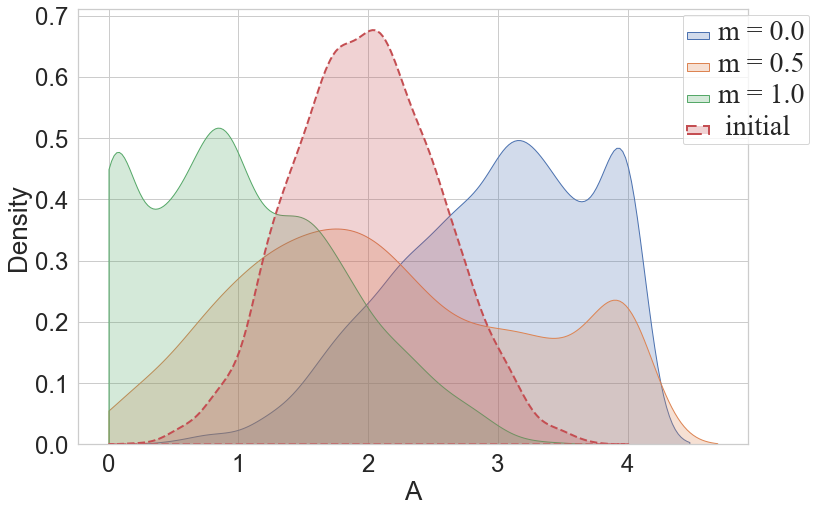}
    \label{fig:4}

    }
    \hspace{-2mm}
    \subfigure[XL]
    {
    \includegraphics[scale=0.27]{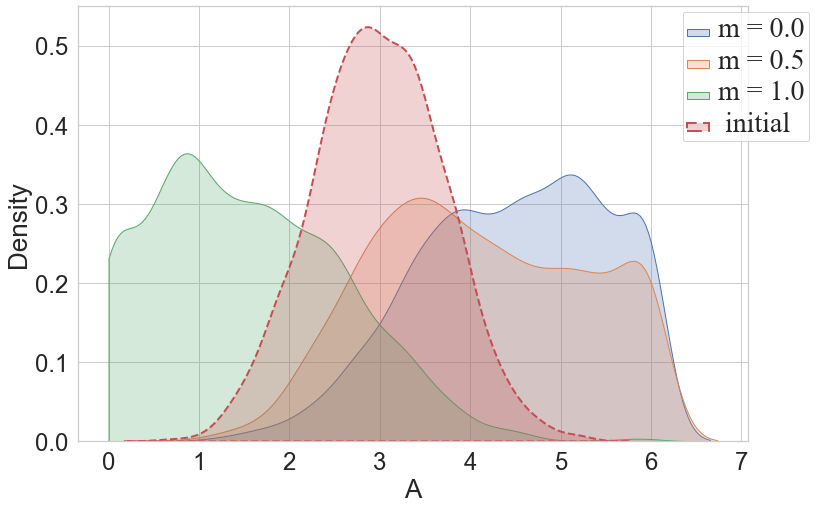}
    \label{fig:6}

    }
    \hspace{-2mm}
    \subfigure[WS]
    {
    \includegraphics[scale=0.27]{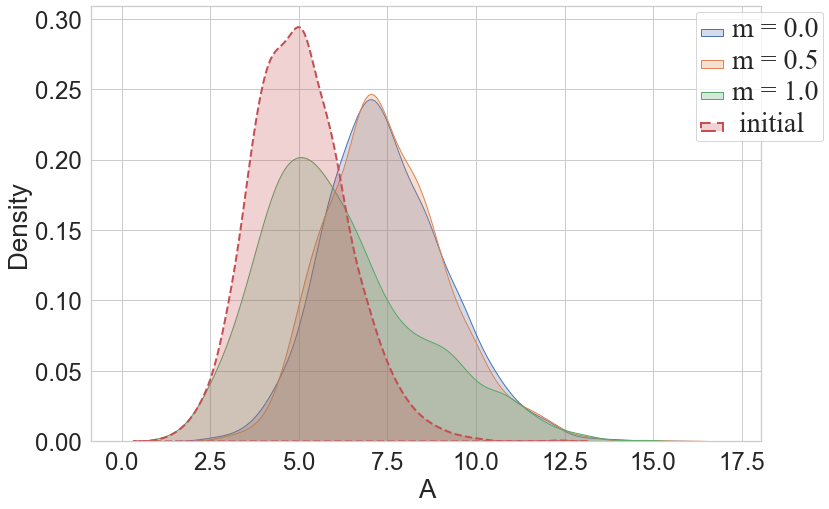}
    \label{fig:WS}

    }
    \hspace{-2mm}
    \subfigure[NW]
    {
    \includegraphics[scale=0.27]{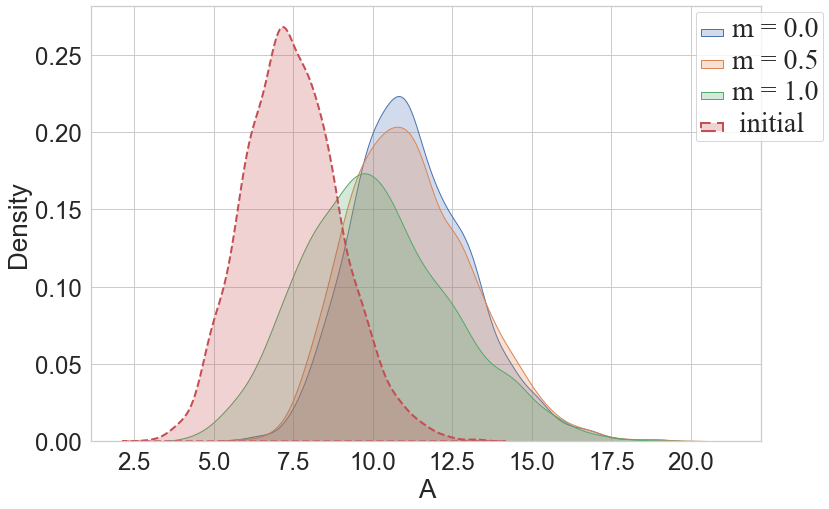}
    \label{fig:NW}

    }
    \caption{\textbf{The probability density distribution of relationship index in the initial and steady states under various parameter settings.} Different panels display the starting and final distribution of relationship index for different graphs, as HL (a), SL (b), XL (c), WS (d), and NW (e). For all cases we used $p = 0.9$ and $b=1.5$ parameters. Horizontal axis shows the actual vales of individual relationship index, while vertical axis shows the probability density of $A$. In all panels the initial normal distribution is shown by dashed red line, while the final distributions for different $m$ values are shown by blue, orange, and green as indicated in the legends.}
    \label{fig6}
\end{figure*}

\subsection{Coevolution of relationship index}
In order to obtain a deeper understanding of the overall and localized characteristics of player relationships, we computed the average centrality values of all agents within each network under varying parameters conditions. Furthermore, we analyzed the centrality distributions for all agents across different parameter combinations, encompassing both their initial and final states. These investigations allowed us to uncover insights into how different network structures and parameter configurations affect the evolution of cooperative behaviors among competing agents.

First, we measured the average relationship index of agents in dependence of the fitness coupling parameter $m$ for all networks. Fig.~\ref{fig5} summarizes our results where we chose three specific $b=1.0$, 1.5 and 2.0 values. They represent low, medium and high temptation. As a robust feature, we can observe that the average relationship index decreases with an increase in parameter $m$. This is valid independently of the temptation value. However, a higher $b$ value leads to a more pronounced and rapid decline in the mean of relationship index. For the lattice topologies (HL, SL and XL) the average relationship index exhibits an initial decline followed by a stabilization trend in the large $m$ region. Notably, within this range, the mean of relationship index on WS consistently demonstrates a decreasing trend. In particular, for both HL and SL, regardless of the chosen $b$, as $m$ becomes sufficiently large, the average relationship index eventually converges to a common value. Evidently, owing to variations in the average degree of the four network types, the average relationship index for all agents increases as the average network degree rises. Accordingly, agents on WS exhibit the highest average relationship index, reaching approximately 7.5, while the maximum for HL is around 2.2. Additionally, it is worth noting that the average relationship index in WS consistently remains higher than the values for other three network types.

Beside the average of $\bar{A}$ we also measured the distribution of relationship index for all networks, shown in different panels of Fig.~6. To evaluate the final stationary distributions correctly we also plot the initial normal distributions. For a proper comparison we used the same  $b=1.5$, $p=0.9$ values for three representative $m=0$, 0.5, and 1.0 values of coupling weight of fitness function. The initial distribution of the relationship index across all networks closely approximates a normal distribution, which is a straightforward consequence of the uniform starting distribution of $W$ values. When analyzing the interplay between $f_c$ and the relationship index distributions, we observe a distinct trend for HL, SL, and XL cases. When $f_c$ approaches or equals 1, the relationship index distribution exhibits a noticeable left-skew. In the other extreme case, when $f_c$ approaches or equals 0, the relationship index distribution leans towards the right-skew. In other words, when cooperation dominates, most agents tend to have relatively higher relationship index values, whereas in times of cooperation dwindling, the majority of agents' relationship index tend to converge toward 0. Under these parameter conditions, both the distribution of relationship index for HL and SL in the stationary state exhibit two distinct prominent peaks. In contrast, the final relationship index distribution in the HL appears relatively stable and lacks distinct peaks. In the context of our research parameter space, it is noteworthy that there is no pronounced skewness in the distribution of relationship index for the WS graph. To check the robustness of our findings obtained for random graph, beside WS network we also applied Newman-Watts (NW) network. Their comparison how relationship index distribution evolves can be seen in Fig.~\ref{fig6}. As panels~(d) and (e) indicate, there is a strong similarity for $m=0$ and 0.5, while there is a slight difference for the extreme case of $m=1$. More importantly, for both random graphs the initial and final relationship index distributions practically approximate the normal distribution.

\begin{table*}

\caption{\label{tab} \textbf{Kurtosis and skewness of the relationship index distribution for five different networks.}
The initial distribution of relationship index of HL, SL, and XL exhibits negative kurtosis and skewness values close to 0. The skewness experiences significant alterations by increasing $m$, which notably affects the $f_c$ value. In contrast, WS and NW networks show positive initial kurtosis and skewness values close to 0. Despite variations in $m$, the skewness remains relatively constant, while a significant decrease in kurtosis is observed when $m$ reaches 1.0.}

\begin{ruledtabular}
\begin{tabular}{ccccccccc}
 &\multicolumn{4}{c}{Kurtosis}&\multicolumn{4}{c}{Skewness}\\
 Networks&Initial&$m=0.0$&$m=0.5$&$m=1.0$&Initial&$m=0.0$&$m=0.5$&$m=1.0$\\ \hline
 HL&-0.336&-0.267&0.601&-0.505&-0.016&-0.589&0.863&0.541\\
 SL&-0.275&-0.423&-0.952&-0.521&0.063&-0.499&0.197&0.463\\
 XL&-0.156&-0.556&0.010&0.532&0.020&-0.387&-0.958&-0.109\\
 WS&0.486&0.404&0.617&0.214&0.115&0.408&0.527&0.716\\
 NW&0.275&0.421&0.267&-0.027&0.067&0.478&0.513&0.392\\
\end{tabular}
\end{ruledtabular}

\end{table*}

In order to quantify the distributions of relationship index for different networks we summarize the kurtosis and skewness values both for initial and final states obtained for $m=0, 0.5$ and 1. The results are summarized in Table~1. Notably, kurtosis measures the sharpness or flatness of the distribution, while skewness indicates its asymmetry. The initial distribution of relationship index in regular lattices (HL, SL, and XL)  exhibits negative kurtosis, while random small-world networks (WS and NW)  show positive initial kurtosis. Besides, the skewness values for all networks are close to 0 in the initial distribution. More precisely, the distribution of HL initially exhibits negative kurtosis, suggesting a relatively flat shape. However, at $m = 0.5$, the kurtosis increases to 0.601, indicating a transition towards a more peaked distribution. Skewness undergoes substantial changes, influenced by the change of $m$, particularly notable between $m = 0.0$ and $m = 0.5$, indicating sensitivity to change in cooperation. In agreement to the graphical observations, the distribution of relationship index in SL is most flattened when $m$ is set to 0.5, as evidenced by a kurtosis value of -0.952. Simultaneously, the skewness is close to zero (0.19), indicating a transition to a more symmetrical distribution. As $m$ increases, kurtosis of XL becomes positive, indicating a transition toward a more peaked distribution. A noteworthy decrease in skewness at $m = 0.5$, reducing to -0.958, signifies a pronounced shift towards a more negatively skewed distribution. Remarkably, WS network shows a peaked distribution with positive kurtosis across different $m$ values, and skewness remains relatively constant, suggesting minimal asymmetry variation in response to changes in the cooperation level. For NW network, a subtle decrease in kurtosis is at $m = 1.0$ indicating a shift towards a less peaked distribution, and skewness remains stable and positive.

\section{Conclusion and outlook}
\label{sec:conclusion}
The main scope of our study was to reveal how strategy choice and the intensity of relationship form each other when individual interest of a player is in conflict with collective goal. Accordingly, the strength of links between neighbors is time dependent, which is affected by the actual strategy of involved partners. To underline the importance of relationship we assumed that the fitness, which drives the strategy imitation process, depends not only on the payoff values determined by actual strategies, but also on the so-called individual relationship index which is a time accumulated fruit of the history with neighbors. The above described coupling can be well studied in the framework of a coevolutionary multiplex system where we distinguish a relationship and interaction layers. 
 
In the relationship layer there are declared connection between neighbors, but the intensity of their relation is case specific. More precisely, while this $W$ weight is always between 0 and 1, its value is increased with an $\epsilon$ when neighbors both cooperate, but is decreased by the same amount for mutual defection. The actual $W$ value expresses the player's willingness to interact with the partner which happens with probability $p$. Otherwise, a randomly selected players are chosen to play with probability $1-p$. Owing to diversity in network structures and the strength of interpersonal relationships, every agent possesses an individualized relationship index and in combination with the individual payoff they determine collectively an agent's fitness. Naturally, the latter value plays a decisive role in strategy imitation process. The basic conflict is characterized by weak prisoner's dilemma game and further key model parameters are $p$, the link between relationship and interaction layers, and $m$ which expresses the relative importance of relationship index in the extended fitness function. To gain a comprehensive view about the potential system behavior we have tested different topologies including honeycomb, square, hexagonal lattices, Watts-Strogatz and Newman-Watts small-world networks. 

There are several generally valid observations. As anticipated, the global cooperation level diminishes with increasing temptation, although the decay of this function may exhibit substantial variations based on the network topology. There is a gradual decline for honeycomb, hexagonal, and small-world network, while the change is sudden for square lattice. The latter topology plays a borderline between the small-degree and large-degree graphs that explains this specific behavior. In a contra-intuitive way, the coupling of payoff and relationship index in the extended fitness function resulted in a lower cooperation level. This behavior can be valid for all graph structures, but it is more spectacular when the average degree is small in the graph. It simply means that by enforcing the fitness with an index summarizing the cumulative history with neighbors does not support, but hinders the evolution of cooperation. This surprising phenomenon can be explained by the fact that a pure payoff based fitness function provides a clean feedback between the strategy change and actual environment. To consider the cumulative history of partnership, however, mitigates this coupling which is enjoyed by defector strategy better.
 
We also detected a clear connection between the cooperation level and the distribution of relationship index. In honeycomb, square, and hexagonal lattices a dominance of cooperators results in a left-skewed distribution, while a dominance of defectors leads to a right-skewed distribution. Interestingly, this pattern is not fully noticed in WS or NW networks where relatively large degree generally hinders the decay of cooperation, hence only the shift toward larger relationship values is detected.

This study adopts multiplex networks to investigate the phenomenon where agents in real-life often establish relationships before engaging in game interactions. Meanwhile, there are still some shortcomings in our model that can be improved. For example, we used the simplest linear coupling in the fitness calculation formula and alternative versions can also be justified. Furthermore, we only considered the change of weight of links in the relationship layer without considering proper break and building alternative connections. Other extensions that can also be explored to gain deeper insights into the spatial evolution of cooperation under stochastic risks. For instance, agents who have stable relationships may exhibit varying probabilities of engaging in strategic interactions in the interaction layer, which can be effectively modeled by using random variables. 

\section*{Acknowledgements}
This research was supported by the National Nature Science Foundation of China (NSFC) under Grant No.62206230, the Humanities and Social Science Fund of Ministry of Education of the People's Republic of China under Grant 21YJCZH028, the Natural Science Foundation of Chongqing under Grant No. CSTB2023NSCQ-MSX0064, and the National Research, Development and Innovation Office (NKFIH) under Grant No. K142948.

%\bibliography{aipsamp}

\end{document}